\documentclass{article}
\usepackage{amsmath}

 \evensidemargin0in \oddsidemargin0in \topmargin10pt
\begin{document}

\title{Relation between Optical Fresnel transformation and quantum
tomography in two-mode entangled case\thanks{{\small Work supported by the
President Foundation of Chinese Academy of Science and the National Natural
Science Foundation of China (Grant No 10874174), and the Research Foundation
of the Education Department of Jiangxi Province.}}}
\author{Hong-yi Fan$^{1}$ and Li-yun Hu$^{2}$\thanks{{\small Corresponding
author. Tel./fax: +86 7918120370. E-mail address: hlyun2008@126.com.}} \\
$^{1}${\small Department of Material Science and Engineering, }\\
{\small University of Science and Technology of China, Hefei, Anhui 230026,
China}\\
$^{2}${\small College of Physics \& Communication Electronics, Jiangxi
Normal University, Nanchang 330022, China}}
\maketitle

\begin{abstract}
{\small Similar in spirit to the preceding work }[{\small Opt. Commun. 282
(2009) 3734] where the relation} {\small between} {\small optical Fresnel
transformation and quantum tomography is revealed}, {\small we study this
kind of relationship in the two-mode entangled case. We show that under the
two-mode Fresnel transformation the bipartite entangled state density }$%
\left\vert \eta \right\rangle \left\langle \eta \right\vert ${\small \
becomes density operator }$F_{2}\left\vert \eta \right\rangle \left\langle
\eta \right\vert F_{2}^{\dagger }=\left\vert \eta \right\rangle
_{s,rs,r}\left\langle \eta \right\vert ${\small , which is just the Radon
transform of the two-mode Wigner operator }$\Delta \left( \sigma ,\gamma
\right) ${\small \ in entangled form, i.e.,}%
\begin{equation*}
\left\vert \eta \right\rangle _{s,rs,r}\left\langle \eta \right\vert =\pi
\int d^{2}\gamma d^{2}\sigma \delta \left( \eta _{2}-D\sigma _{2}+B\gamma
_{1}\right) \delta \left( \eta _{1}-D\sigma _{1}-B\gamma _{2}\right) \Delta
\left( \sigma ,\gamma \right) ,
\end{equation*}%
{\small where }$F_{2}${\small \ is an two-mode Fresnel operator in quantum
optics, and }$s,r${\small \ are the complex-value expression of }${\small (}%
A,B,C,D).$ {\small So the probability distribution for the Fresnel
quadrature phase is the tomography (Radon transform of the two-mode Wigner
function), correspondingly, }$_{s,r}\left\langle \eta \right\vert \left.
\psi \right\rangle =\left\langle \eta \right\vert F_{2}^{\dagger }\left\vert
\psi \right\rangle .${\small \ Similarly, we find }%
\begin{equation*}
F_{2}\left\vert \xi \right\rangle \left\langle \xi \right\vert
F_{2}^{\dagger }=\left\vert \xi \right\rangle _{s,rs,r}\left\langle \xi
\right\vert =\pi \int d^{2}\sigma d^{2}\gamma \delta \left( \xi _{1}-A\sigma
_{1}-C\gamma _{2}\right) \delta \left( \xi _{2}-A\sigma _{2}+C\gamma
_{1}\right) \Delta \left( \sigma ,\gamma \right) ,
\end{equation*}%
{\small where }$\left\vert \xi \right\rangle ${\small \ is the conjugated
state to }$\left\vert \eta \right\rangle ${\small .}
\end{abstract}

\section{Introduction}

In Ref.\cite{r1}, by using the technique of integration within an ordered
product (IWOP) of operators and the coherent state representation \cite%
{r2,r3} we have proved that corresponding to optical Fresnel transformation
characteristic of ray transfer matrix elements $(A,B,C,D),$ $AD-BC=1$,
connecting the input light field $f\left( x\right) $ and output light field $%
g\left( x^{\prime }\right) $ by Fresnel integration \cite{r4,r5,r6}
\begin{equation}
g\left( x^{\prime }\right) =\frac{1}{\sqrt{2\pi iB}}\int_{-\infty }^{\infty
}\exp \left[ \frac{i}{2B}\left( Ax^{2}-2x^{\prime }x+Dx^{\prime 2}\right) %
\right] f\left( x\right) dx.  \label{1}
\end{equation}%
$\;$ there exists Fresnel operator $F_{1}(r,s)$\ in quantum optics \cite{r7},%
\begin{eqnarray}
F_{1}\left( r,s\right) &=&\sqrt{s}\int \frac{d^{2}z}{\pi }\left\vert
sz-rz^{\ast }\right\rangle \left\langle z\right\vert  \notag \\
&=&\frac{1}{\sqrt{s^{\ast }}}\exp \left( -\frac{ra^{\dagger 2}}{2s^{\ast }}%
\right) \colon \exp \left\{ \left( \frac{1}{s^{\ast }}-1\right) a^{\dagger
}a\right\} \colon \exp \left( \frac{r^{\ast }a^{2}}{2s^{\ast }}\right) ,\;\;
\label{2}
\end{eqnarray}%
such that in the coordinate $\left\langle x\right\vert $ representation\cite%
{r1,r8}
\begin{equation}
\left\langle x^{\prime }\right\vert F_{1}\left( r,s\right) \left\vert
x\right\rangle =\frac{1}{\sqrt{2\pi iB}}\exp \left[ \frac{i}{2B}\left(
Ax^{2}-2x^{\prime }x+Dx^{\prime 2}\right) \right] ,  \label{4}
\end{equation}%
where $(s,r),$ $\left\vert s\right\vert ^{2}-\left\vert r\right\vert ^{2}=1$%
, are related to a classical ray transfer matrix $\left(
\begin{array}{cc}
A & B \\
C & D%
\end{array}%
\right) $ by%
\begin{equation}
s=\frac{1}{2}\left[ A+D-i\left( B-C\right) \right] ,\;r=-\frac{1}{2}\left[
A-D+i\left( B+C\right) \right] ,  \label{3}
\end{equation}%
the unimodularity condition $AD-BC=1$ is equivalent to $\left\vert
s\right\vert ^{2}-\left\vert r\right\vert ^{2}=1.$ If we let $f\left(
x\right) =\left\langle x\right\vert \left. f\right\rangle $, then Eq.(\ref{1}%
) is expressed as%
\begin{equation}
g\left( x^{\prime }\right) =\int_{-\infty }^{\infty }\left\langle x^{\prime
}\right\vert F_{1}\left( r,s\right) \left\vert x\right\rangle \left\langle
x\right\vert \left. f\right\rangle dx=\left\langle x^{\prime }\right\vert
F_{1}\left( r,s\right) \left\vert f\right\rangle ,  \label{5}
\end{equation}%
which is just the quantum mechanical version of Fresnel
transformation.

In a preceding paper \cite{r9}, we also found that under the Fresnel
transformation the pure position density $\left\vert x\right\rangle
\left\langle x\right\vert $\ becomes the tomographic density $\left\vert
x\right\rangle _{s,rs,r}\left\langle x\right\vert $, which is just the Radon
transform of the Wigner operator $\Delta \left( x,p\right) ,$ i.e.,%
\begin{equation}
F_{1}\left\vert x\right\rangle \left\langle x\right\vert F_{1}^{\dagger
}=\left\vert x\right\rangle _{s,rs,r}\left\langle x\right\vert
=\int_{-\infty }^{\infty }dp^{\prime }dx^{\prime }\delta \left[ x-\left(
Dx^{\prime }-Bp^{\prime }\right) \right] \Delta \left( x^{\prime },p^{\prime
}\right) .  \label{6}
\end{equation}%
So the probability distribution for the Fresnel quadrature phase is the
tomography (Radon transform of Wigner function \cite{r9a,r9b,r9c}), and the
tomogram of a state $\left\vert \psi \right\rangle $\ is just the wave
function of its Fresnel transformed state $F_{1}^{\dagger }\left\vert \psi
\right\rangle ,$\ i.e. $_{s,r}\left\langle x\right\vert \left. \psi
\right\rangle =\left\langle x\right\vert F_{1}^{\dagger }\left\vert \psi
\right\rangle ,$ and $\left\vert x\right\rangle _{s,r}$ is expressed as%
\begin{equation}
\left\vert x\right\rangle _{s,r}=\frac{\pi ^{-1/4}}{\sqrt{D+iB}}\exp \left\{
-\frac{A-iC}{D+iB}\frac{x^{2}}{2}+\frac{\sqrt{2}x}{D+iB}a^{\dagger }-\frac{%
D-iB}{D+iB}\frac{a^{\dagger 2}}{2}\right\} \left\vert 0\right\rangle .
\label{7}
\end{equation}%
In this Communication we want to generalize the above conclusion to two-mode
entangled case. Firstly, we extend Eq. (\ref{4}) to the two-dimensional
Fresnel transformation,
\begin{equation}
\mathcal{K}_{2}\left( \eta ^{\prime },\eta \right) =\frac{1}{2iB\pi }\exp %
\left[ \frac{i}{2B}(A\left\vert \eta \right\vert ^{2}-\left( \eta \eta
^{\prime \ast }+\eta ^{\ast }\eta ^{\prime }\right) +D\left\vert \eta
^{\prime }\right\vert ^{2})\right] ,  \label{8}
\end{equation}%
where $\eta $ is a complex number, then we construct the two-mode Fresnel
operator $F_{2}\left( r,s\right) $ such that its transformation matrix
element in the entangled state $\left\vert \eta \right\rangle $
representation (see below Eq.(\ref{16})) is just the two-dimensional Fresnel
transformation, i.e., $\mathcal{K}_{2}^{\left( r,s\right) }\left( \eta
^{\prime },\eta \right) =\frac{1}{\pi }\left\langle \eta ^{\prime
}\right\vert F_{2}\left( r,s\right) \left\vert \eta \right\rangle ,$ then we
shall prove
\begin{equation}
F_{2}\left\vert \eta \right\rangle \left\langle \eta \right\vert
F_{2}^{\dagger }=\left\vert \eta \right\rangle _{s,rs,r}\left\langle \eta
\right\vert =\pi \int d^{2}\gamma d^{2}\sigma \delta \left( \eta
_{2}-D\sigma _{2}+B\gamma _{1}\right) \delta \left( \eta _{1}-D\sigma
_{1}-B\gamma _{2}\right) \Delta \left( \sigma ,\gamma \right) ,  \label{9}
\end{equation}%
i.e., we show that $\left\vert \eta \right\rangle _{s,rs,r}\left\langle \eta
\right\vert $ is just the Radon transform of the entangled Wigner operator $%
\Delta \left( \sigma ,\gamma \right) $

Our paper is arranged as follows. In section 2, we briefly review the
two-mode Fresnel operator $F_{2}\left( r,s\right) $ and then derive the 2D
Fresnel transformation in entangled state $\left\vert \eta \right\rangle $
representation and introduce a new representation $\left\vert \eta
\right\rangle _{s,r}(=F_{2}\left( r,s\right) \left\vert \eta \right\rangle )$
in section 3. Section 4 is devoted to proving Eq.(\ref{9}), i.e., $%
\left\vert \eta \right\rangle _{s,r\text{ }s,r}\left\langle \eta \right\vert
$ as Radon transform of entangled Wigner operator. Similar discussions are
moved to the Fresnel transformation in its `frequency domain' in section 5.
In the last section, we derive the inverse Radon transformation of entangled
Wigner operator.

\section{Two-mode Fresnel operator}

Similar in spirit to the single-mode case, we introduce the two-mode Fresnel
operator $F_{2}\left( r,s\right) $ through the following 2-mode coherent
state representation \cite{r1,r10}, i.e.,
\begin{equation}
F_{2}\left( r,s\right) =s\int \frac{d^{2}z_{1}d^{2}z_{2}}{\pi ^{2}}%
\left\vert sz_{1}+rz_{2}^{\ast },rz_{1}^{\ast }+sz_{2}\right\rangle
\left\langle z_{1},z_{2}\right\vert ,  \label{10}
\end{equation}%
which indicates that $F_{2}\left( r,s\right) $ is a mapping of classical
symplectic transform $\left( z_{1},z_{2}\right) \rightarrow \left(
sz_{1}+rz_{2}^{\ast },rz_{1}^{\ast }+sz_{2}\right) $ in phase space, where $%
\left\vert z_{1},z_{2}\right\rangle =\exp \left\{ -\frac{1}{2}\left\vert
z_{1}\right\vert ^{2}-\frac{1}{2}\left\vert z_{2}\right\vert
^{2}+z_{1}a_{1}^{\dag }+z_{1}a_{1}^{\dag }\right\} \left\vert
00\right\rangle $ is a usual two-mode coherent state. Concretely, the ket in
(\ref{10}) is
\begin{equation}
\left\vert sz_{1}+rz_{2}^{\ast },rz_{1}^{\ast }+sz_{2}\right\rangle \equiv
\left\vert sz_{1}+rz_{2}^{\ast }\right\rangle _{1}\otimes \left\vert
rz_{1}^{\ast }+sz_{2}\right\rangle _{2},\text{ }  \label{11}
\end{equation}%
$s$ and $r$ are complex and satisfy the unimodularity condition $\left\vert
s\right\vert ^{2}-\left\vert r\right\vert ^{2}=1$. Using the IWOP technique
\cite{r11,r12} and the normal ordering of the vacuum projector $\left\vert
00\right\rangle \left\langle 00\right\vert =\colon \exp \left(
-a_{1}^{\dagger }a_{1}-a_{2}^{\dagger }a_{2}\right) \colon ,$ we perform the
integral in (\ref{10}) and obtain%
\begin{align}
F_{2}\left( r,s\right) & =s\int \frac{1}{\pi ^{2}}d^{2}z_{1}d^{2}z_{2}\colon
\exp [-|s|^{2}\left( |z_{1}|^{2}+|z_{2}|^{2}\right) -r^{\ast
}sz_{1}z_{2}-rs^{\ast }z_{1}^{\ast }z_{2}^{\ast }  \notag \\
& +\left( sz_{1}+rz_{2}^{\ast }\right) a_{1}^{\dagger }+\left( rz_{1}^{\ast
}+sz_{2}\right) a_{2}^{\dagger }+z_{1}^{\ast }a_{1}+z_{2}^{\ast
}a_{2}-a_{1}^{\dagger }a_{1}-a_{2}^{\dagger }a_{2}]\colon   \notag \\
& =\frac{1}{s^{\ast }}\exp \left( \frac{r}{s^{\ast }}a_{1}^{\dagger
}a_{2}^{\dagger }\right) \colon \exp \left[ \left( \frac{1}{s^{\ast }}%
-1\right) \left( a_{1}^{\dagger }a_{1}+a_{2}^{\dagger }a_{2}\right) \right]
\colon \exp \left( -\frac{r^{\ast }}{s^{\ast }}a_{1}a_{2}\right)   \notag \\
& =\exp \left( \frac{r}{s^{\ast }}a_{1}^{\dagger }a_{2}^{\dagger }\right)
\exp [\left( a_{1}^{\dagger }a_{1}+a_{2}^{\dagger }a_{2}+1\right) \ln \left(
s^{\ast }\right) ^{-1}]\exp \left( -\frac{r^{\ast }}{s^{\ast }}%
a_{1}a_{2}\right) .  \label{12}
\end{align}%
Thus $F_{2}\left( r,s\right) $ induces the transforms
\begin{equation}
F_{2}\left( r,s\right) a_{1}F_{2}^{-1}\left( r,s\right) =s^{\ast
}a_{1}-ra_{2}^{\dagger },\text{ \ }F_{2}\left( r,s\right)
a_{2}F_{2}^{-1}\left( r,s\right) =s^{\ast }a_{2}-ra_{1}^{\dagger },
\label{13}
\end{equation}%
and $F_{2}$ is actually a generalized 2-mode squeezing operator \cite%
{r12a,r12b}.

$F_{2}\left( r,s\right) $ abides by the group multiplication rule. Using the
IWOP technique and (\ref{10}) we obtain%
\begin{align}
& F_{2}\left( r,s\right) F_{2}\left( r^{\prime },s^{\prime }\right)  \notag
\\
& =ss^{\prime }\int \frac{d^{2}z_{1}d^{2}z_{2}d^{2}z_{1}^{\prime
}d^{2}z_{2}^{\prime }}{\pi ^{4}}\colon \exp \{-|s|^{2}\left(
|z_{1}|^{2}+|z_{2}|^{2}\right) -r^{\ast }sz_{1}z_{2}  \notag \\
& -rs^{\ast }z_{1}^{\ast }z_{2}^{\ast }-\frac{1}{2}[|z_{1}^{\prime
}|^{2}+|z_{2}^{\prime }|^{2}+|s^{\prime }z_{1}^{\prime }+r^{\prime
}z_{2}^{\prime \ast }|^{2}+|r^{\prime }z_{1}^{\prime \ast }+s^{\prime
}z_{2}^{\prime }|^{2}]  \notag \\
& +\left( sz_{1}+rz_{2}^{\ast }\right) a_{1}^{\dagger }+\left( rz_{1}^{\ast
}+sz_{2}\right) a_{2}^{\dagger }+z_{1}^{\prime \ast }a_{1}+z_{2}^{\prime
\ast }a_{2}  \notag \\
& +z_{1}^{\ast }\left( s^{\prime }z_{1}^{\prime }+r^{\prime }z_{2}^{\prime
\ast }\right) +z_{2}^{\ast }\left( r^{\prime }z_{1}^{\prime \ast }+s^{\prime
}z_{2}^{\prime }\right) -a_{1}^{\dagger }a_{1}-a_{2}^{\dagger }a_{2}\}\colon
\notag \\
& =\frac{1}{s^{\prime \prime \ast }}\exp \left( \frac{r^{\prime \prime }}{%
2s^{\prime \prime \ast }}a_{1}^{\dagger }a_{2}^{\dagger }\right) \colon \exp
\left\{ \left( \frac{1}{s^{\prime \prime \ast }}-1\right) \left(
a_{1}^{\dagger }a_{1}+a_{2}^{\dagger }a_{2}\right) \right\} \colon \exp
\left( -\frac{r^{\prime \prime \ast }}{2s^{\prime \prime \ast }}%
a_{1}a_{2}\right)  \notag \\
& =F_{2}\left( r^{\prime \prime },s^{\prime \prime }\right) ,  \label{14}
\end{align}%
where $\left( r^{\prime \prime },s^{\prime \prime }\right) $ are given by
\begin{equation}
\left(
\begin{array}{cc}
s^{\prime \prime } & -r^{\prime \prime } \\
-r^{\ast \prime \prime } & s^{\ast \prime \prime }%
\end{array}%
\right) =\left(
\begin{array}{cc}
s & -r \\
-r^{\ast } & s^{\ast }%
\end{array}%
\right) \left(
\begin{array}{cc}
s^{\prime } & -r^{\prime } \\
-r^{\prime \ast } & s^{\prime \ast }%
\end{array}%
\right) .  \label{15}
\end{equation}%
Therefore, Eq.(\ref{14}) is a loyal representation of the multiplication
rule for ray transfer matrices in the sense of \textit{Matrix Optics}.$\ $

\section{Two-mode Fresnel transformation in entangled state representations}

By introducing the bipartite entangled state $\left\vert \eta \right\rangle $
\cite{r13,r14}
\begin{equation}
\left\vert \eta \right\rangle =\exp \left[ -\frac{1}{2}\left\vert \eta
\right\vert ^{2}+\eta a_{1}^{\dagger }-\eta ^{\ast }a_{2}^{\dagger
}+a_{1}^{\dagger }a_{2}^{\dagger }\right] \left\vert 00\right\rangle ,
\label{16}
\end{equation}%
$\left\vert \eta =\eta _{1}+i\eta _{2}\right\rangle $ is the common
eigenstate of relative coordinate $Q_{1}-Q_{2}$ and the total momentum $%
P_{1}+P_{2}$, i.e.,
\begin{equation}
\left( Q_{1}-Q_{2}\right) \left\vert \eta \right\rangle =\sqrt{2}\eta
_{1}\left\vert \eta \right\rangle ,\,\ \left( P_{1}+P_{2}\right) \left\vert
\eta \right\rangle =\sqrt{2}\eta _{2}\left\vert \eta \right\rangle ,
\label{17}
\end{equation}%
where $Q_{i}=\left( a_{i}+a_{i}^{\dagger }\right) /\sqrt{2},\ P_{i}=\left(
a_{i}-a_{i}^{\dagger }\right) /(\mathtt{i}\sqrt{2}),$ $\left( i=1,2.\right) ,
$ are coordinate and momentum operators, respectively. $\left\vert \eta
\right\rangle $ compose a complete set $\int \frac{d^{2}\eta }{\pi }%
\left\vert \eta \right\rangle \left\langle \eta \right\vert =1,$
then using the over-completeness relation of the coherent state and
\begin{equation}
\left\langle z_{1},z_{2}\right\vert \left. \eta \right\rangle =\exp \left[ -%
\frac{1}{2}(\left\vert z_{1}\right\vert ^{2}+\left\vert z_{2}\right\vert
^{2}+\left\vert \eta \right\vert ^{2})+\eta z_{1}^{\ast }-\eta ^{\ast
}z_{2}^{\ast }+z_{1}^{\ast }z_{2}^{\ast }\right] ,  \label{18}
\end{equation}%
as well as
\begin{align}
\left\langle z_{1}^{\prime },z_{2}^{\prime }\right\vert F_{2}\left(
r,s\right) \left\vert z_{1},z_{2}\right\rangle & =\frac{1}{s^{\ast }}\exp
\left\{ -\frac{1}{2}(\left\vert z_{1}\right\vert ^{2}+\left\vert
z_{2}\right\vert ^{2}+\left\vert z_{1}^{\prime }\right\vert ^{2}+\left\vert
z_{2}^{\prime }\right\vert ^{2})\right.   \notag \\
& \left. +\frac{r}{s^{\ast }}z_{1}^{\prime \ast }z_{2}^{\prime \ast }-\frac{%
r^{\ast }}{s^{\ast }}z_{1}z_{2}+\frac{1}{s^{\ast }}\left( z_{1}^{\prime \ast
}z_{1}+z_{2}^{\prime \ast }z_{2}\right) \right\} ,  \label{19}
\end{align}%
we can calculate the integral kernel%
\begin{align}
\mathcal{K}_{2}^{\left( r,s\right) }\left( \eta ^{\prime },\eta \right) & =%
\frac{1}{\pi }\left\langle \eta ^{\prime }\right\vert F_{2}\left( r,s\right)
\left\vert \eta \right\rangle   \notag \\
& =\int \frac{d^{2}z_{1}d^{2}z_{2}d^{2}z_{1}^{\prime }d^{2}z_{2}^{\prime }}{%
\pi ^{5}}\left\langle \eta ^{\prime }\right\vert \left. z_{1}^{\prime
},z_{2}^{\prime }\right\rangle \left\langle z_{1}^{\prime },z_{2}^{\prime
}\right\vert F_{2}\left( r,s\right) \left\vert z_{1},z_{2}\right\rangle
\left\langle z_{1},z_{2}\right. \left\vert \eta \right\rangle   \notag \\
& =\frac{1}{s^{\ast }}\int \frac{d^{2}z_{1}d^{2}z_{2}d^{2}z_{1}^{\prime
}d^{2}z_{2}^{\prime }}{\pi ^{5}}\exp \left[ -(\left\vert z_{1}\right\vert
^{2}+\left\vert z_{2}\right\vert ^{2}+\left\vert z_{1}^{\prime }\right\vert
^{2}+\left\vert z_{2}^{\prime }\right\vert ^{2})-\frac{1}{2}(\left\vert \eta
^{\prime }\right\vert ^{2}+\left\vert \eta \right\vert ^{2})\right]   \notag
\\
& \times \exp \left[ z_{1}^{\ast }z_{2}^{\ast }+\eta z_{1}^{\ast }-\frac{%
r^{\ast }}{s^{\ast }}z_{1}z_{2}+\frac{z_{1}^{\prime \ast
}z_{1}+z_{2}^{\prime \ast }z_{2}}{s^{\ast }}+\frac{r}{s^{\ast }}%
z_{1}^{\prime \ast }z_{2}^{\prime \ast }+z_{1}^{\prime }z_{2}^{\prime }+\eta
^{\prime \ast }z_{1}^{\prime }-\eta ^{\prime }z_{2}^{\prime }-\eta ^{\ast
}z_{2}^{\ast }\right]   \notag \\
& =\frac{1}{\left( r^{\ast }+s^{\ast }-r-s\right) \pi }\exp \left[ \frac{%
\left( r^{\ast }-s\right) \left\vert \eta \right\vert ^{2}-\left( r+s\right)
\left\vert \eta ^{\prime }\right\vert ^{2}+\eta \eta ^{\prime \ast }+\eta
^{\ast }\eta ^{\prime }}{r^{\ast }+s^{\ast }-r-s}-\frac{\left\vert \eta
^{\prime }\right\vert ^{2}+\left\vert \eta \right\vert ^{2}}{2}\right] .
\label{20}
\end{align}%
Using the relation between $s,r$ and $\left( A,B,C,D\right) $ in Eq.(\ref{3}%
) we see that Eq. (\ref{20}) becomes%
\begin{equation}
\mathcal{K}_{2}^{\left( r,s\right) }\left( \eta ^{\prime },\eta \right) =%
\frac{1}{2iB\pi }\exp \left[ \frac{i}{2B}\left( A\left\vert \eta \right\vert
^{2}-\left( \eta \eta ^{\prime \ast }+\eta ^{\ast }\eta ^{\prime }\right)
+D\left\vert \eta ^{\prime }\right\vert ^{2}\right) \right] \equiv \mathcal{K%
}_{2}^{M}\left( \eta ^{\prime },\eta \right) ,  \label{21}
\end{equation}%
where the superscript $M$ only means the parameters of $\mathcal{K}_{2}^{M}$
are $\left[ A,B;C,D\right] $, and the subscript $2$ implies the
two-dimensional kernel.

Operating $F_{2}\left( r,s\right) $ on $\left\vert \eta \right\rangle $ and
using Eqs.(\ref{12}) and (\ref{18}) yields

\begin{eqnarray}
F_{2}\left( r,s\right) \left\vert \eta \right\rangle  &=&\frac{1}{s^{\ast }}%
\int \frac{d^{2}z_{1}d^{2}z_{2}}{\pi ^{2}}\exp \left[ \frac{r}{s^{\ast }}%
a_{1}^{\dagger }a_{2}^{\dagger }+\left( \frac{1}{s^{\ast }}-1\right) \left(
a_{1}^{\dagger }z_{1}+a_{2}^{\dagger }z_{2}\right) -\frac{r^{\ast }}{s^{\ast
}}z_{1}z_{2}\right] \left\vert z_{1},z_{2}\right\rangle \left\langle
z_{1},z_{2}\right\vert \left. \eta \right\rangle   \notag \\
&=&\frac{1}{s^{\ast }}\int \frac{d^{2}z_{1}d^{2}z_{2}}{\pi ^{2}}\exp \left[
-\left\vert z_{1}\right\vert ^{2}+\frac{1}{s^{\ast }}\left( a_{1}^{\dagger
}-r^{\ast }z_{2}\right) z_{1}+\left( \eta +z_{2}^{\ast }\right) z_{1}^{\ast }%
\right]   \notag \\
&&\times \exp \left[ -\frac{1}{2}\left\vert \eta \right\vert ^{2}-\left\vert
z_{2}\right\vert ^{2}+\frac{1}{s^{\ast }}z_{2}a_{2}^{\dagger }-\eta ^{\ast
}z_{2}^{\ast }+\frac{r}{s^{\ast }}a_{1}^{\dagger }a_{2}^{\dagger }\right]
\left\vert 00\right\rangle   \notag \\
&=&\frac{1}{s^{\ast }}\int \frac{d^{2}z_{2}}{\pi }\exp \left[ -\frac{s^{\ast
}+r^{\ast }}{s^{\ast }}\left\vert z_{2}\right\vert ^{2}+\frac{1}{s^{\ast }}%
\left( a_{2}^{\dagger }-\eta r^{\ast }\right) z_{2}+\frac{1}{s^{\ast }}%
\left( a_{1}^{\dagger }-s^{\ast }\eta ^{\ast }\right) z_{2}^{\ast }\right]
\notag \\
&&\times \exp \left[ +\frac{\eta }{s^{\ast }}a_{1}^{\dagger }+\frac{r}{%
s^{\ast }}a_{1}^{\dagger }a_{2}^{\dagger }-\frac{1}{2}\left\vert \eta
\right\vert ^{2}\right] \left\vert 00\right\rangle   \notag \\
&=&\frac{1}{s^{\ast }+r^{\ast }}\exp \left\{ -\allowbreak \frac{s^{\ast
}-r^{\ast }}{2\left( s^{\ast }+r^{\ast }\right) }\left\vert \eta \right\vert
^{2}+\allowbreak \frac{\eta a_{1}^{\dagger }}{s^{\ast }+r^{\ast }}%
\allowbreak -\allowbreak \frac{\eta ^{\ast }a_{2}^{\dagger }}{s^{\ast
}+r^{\ast }}+\frac{s+r}{s^{\ast }+r^{\ast }}\allowbreak a_{1}^{\dagger
}a_{2}^{\dagger }\right\} \left\vert 00\right\rangle \equiv \left\vert \eta
\right\rangle _{s,r},  \label{22}
\end{eqnarray}%
or
\begin{equation}
\left\vert \eta \right\rangle _{s,r}=\frac{1}{\allowbreak D+iB}\exp \left\{ -%
\frac{\allowbreak A-iC}{2\left( \allowbreak D+iB\right) }\left\vert \eta
\right\vert ^{2}+\frac{\eta a_{1}^{\dagger }}{\allowbreak D+iB}-\frac{\eta
^{\ast }a_{2}^{\dagger }}{\allowbreak D+iB}+\frac{\allowbreak D-iB}{%
\allowbreak D+iB}a_{1}^{\dagger }a_{2}^{\dagger }\right\} \left\vert
00\right\rangle ,  \label{23}
\end{equation}%
where we have used the integration formula%
\begin{equation}
\int \frac{d^{2}z}{\pi }\exp \left( \zeta \left\vert z\right\vert ^{2}+\xi
z+\eta z^{\ast }\right) =-\frac{1}{\zeta }e^{-\frac{\xi \eta }{\zeta }},%
\text{Re}\left( \zeta \right) <0.  \label{24}
\end{equation}%
Noticing the completeness relation and the orthogonality of $\left\vert \eta
\right\rangle $ we immediately derive
\begin{equation}
\int \frac{d^{2}\eta }{\pi }\left\vert \eta \right\rangle
_{s,rs,r}\left\langle \eta \right\vert =1,\text{ }_{s,r}\left\langle \eta
\right\vert \left. \eta ^{\prime }\right\rangle _{s,r}=\pi \delta \left(
\eta -\eta ^{\prime }\right) \delta \left( \eta ^{\ast }-\eta ^{\prime \ast
}\right) ,  \label{25}
\end{equation}%
a generalized entangled state representation $\left\vert \eta \right\rangle
_{s,r}$ with the completeness relation (\ref{25}). From (\ref{23}) we can
see that%
\begin{eqnarray}
a_{1}\left\vert \eta \right\rangle _{s,r} &=&\left( \frac{\eta }{\allowbreak
D+iB}+\frac{\allowbreak D-iB}{\allowbreak D+iB}a_{2}^{\dagger }\right)
\left\vert \eta \right\rangle _{s,r},  \label{26} \\
a_{2}\left\vert \eta \right\rangle _{s,r} &=&\left( -\frac{\eta ^{\ast }}{%
\allowbreak D+iB}+\frac{\allowbreak D-iB}{\allowbreak D+iB}a_{1}^{\dagger
}\right) \left\vert \eta \right\rangle _{s,r},  \label{27}
\end{eqnarray}%
so we have the eigen-equations for $\left\vert \eta \right\rangle _{s,r}$ as
follows%
\begin{eqnarray}
\left[ D\left( Q_{1}-Q_{2}\right) -B\left( P_{1}-P_{2}\right) \right]
\left\vert \eta \right\rangle _{s,r} &=&\sqrt{2}\eta _{1}\left\vert \eta
\right\rangle _{s,r},\text{ }  \label{28} \\
\left[ B\left( Q_{1}+Q_{2}\right) +D\left( P_{1}+P_{2}\right) \right]
\left\vert \eta \right\rangle _{s,r} &=&\sqrt{2}\eta _{2}\left\vert \eta
\right\rangle _{s,r},  \label{29}
\end{eqnarray}%
We can also check Eqs.(\ref{26})-(\ref{29}) by another way (see Appendix).

\section{$\left\vert \protect\eta \right\rangle _{s,r\text{ }%
s,r}\left\langle \protect\eta \right\vert $ as the Radon transform of
entangled Wigner operator}

For two-mode correlated system, it is convenient to express the Wigner
operator in the $\left\vert \eta \right\rangle $ representation as \cite%
{r15,r16,r17,r18}
\begin{equation}
\Delta \left( \sigma ,\gamma \right) =\int \frac{d^{2}\eta }{\pi ^{3}}%
\left\vert \sigma -\eta \right\rangle \left\langle \sigma +\eta \right\vert
e^{\eta \gamma ^{\ast }-\eta ^{\ast }\gamma }.  \label{30}
\end{equation}%
When $\sigma =\alpha -\beta ^{\ast },\;\gamma =\alpha +\beta ^{\ast }$, Eq. (%
\ref{30}) is just equal to the direct product of two single-mode Wigner
operators, i.e., $\Delta \left( \sigma ,\gamma \right) =\Delta \left( \alpha
,\alpha ^{\ast }\right) \otimes \Delta \left( \beta ,\beta ^{\ast }\right) .$
Then according to the Wely correspondence rule \cite{r19}%
\begin{equation}
H\left( a_{1}^{\dagger },a_{2}^{\dagger };a_{1},a_{2}\right) =\int
d^{2}\gamma d^{2}\sigma h\left( \sigma ,\gamma \right) \Delta \left( \sigma
,\gamma \right) ,  \label{31}
\end{equation}%
where $h\left( \sigma ,\gamma \right) $ is the Weyl correspondence of $%
H\left( a_{1}^{\dagger },a_{2}^{\dagger };a_{1},a_{2}\right) ,$ and
\begin{equation}
h\left( \sigma ,\gamma \right) =4\pi ^{2}\mathtt{Tr}\left[ H\left(
a_{1}^{\dagger },a_{2}^{\dagger };a_{1},a_{2}\right) \Delta \left( \sigma
,\gamma \right) \right] ,  \label{32}
\end{equation}%
the classical Weyl correspondence of the projection operator $\left\vert
\eta \right\rangle _{r,sr,s}\left\langle \eta \right\vert $ can be
calculated,
\begin{eqnarray}
&&4\pi ^{2}\mathtt{Tr}\left[ \left\vert \eta \right\rangle
_{r,sr,s}\left\langle \eta \right\vert \Delta \left( \sigma ,\gamma \right) %
\right]   \notag \\
&=&4\pi ^{2}\int \frac{d^{2}\eta ^{\prime }}{\pi ^{3}}\left.
_{r,s}\left\langle \eta \right\vert \left. \sigma -\eta ^{\prime
}\right\rangle \left\langle \sigma +\eta ^{\prime }\right\vert \left. \eta
\right\rangle _{r,s}\right. \exp (\eta ^{\prime }\gamma ^{\ast }-\eta
^{\prime \ast }\gamma )  \notag \\
&=&4\pi ^{2}\int \frac{d^{2}\eta ^{\prime }}{\pi ^{3}}\left\langle \eta
\right\vert F_{2}^{\dagger }\left\vert \sigma -\eta ^{\prime }\right\rangle
\left\langle \sigma +\eta ^{\prime }\right\vert F_{2}\left\vert \eta
\right\rangle \exp (\eta ^{\prime }\gamma ^{\ast }-\eta ^{\prime \ast
}\gamma ).  \label{33}
\end{eqnarray}%
Then using Eq.(\ref{20}) we have%
\begin{equation}
4\pi ^{2}\mathtt{Tr}\left[ \left\vert \eta \right\rangle
_{s,rs,r}\left\langle \eta \right\vert \Delta \left( \sigma ,\gamma \right) %
\right] =\pi \delta \left( \eta _{2}-D\sigma _{2}+B\gamma _{1}\right) \delta
\left( \eta _{1}-D\sigma _{1}-B\gamma _{2}\right) ,  \label{34}
\end{equation}%
which means the following Weyl correspondence
\begin{equation}
\left\vert \eta \right\rangle _{s,rs,r}\left\langle \eta \right\vert =\pi
\int d^{2}\gamma d^{2}\sigma \delta \left( \eta _{2}-D\sigma _{2}+B\gamma
_{1}\right) \delta \left( \eta _{1}-D\sigma _{1}-B\gamma _{2}\right) \Delta
\left( \sigma ,\gamma \right) ,  \label{35}
\end{equation}%
so the projector operator $\left\vert \eta \right\rangle
_{s,rs,r}\left\langle \eta \right\vert $ is just the Radon transformation of
$\Delta \left( \sigma ,\gamma \right) $, $D$ and $B$ are the Radon
transformation parameter. Combining Eqs. (\ref{22})-(\ref{35}) together we
complete the proof (\ref{9}). Therefore, the quantum tomography in two-mode
entangled case is expressed as
\begin{equation}
|_{s,r}\left\langle \eta \right\vert \left. \psi \right\rangle
|^{2}=|\left\langle \eta \right\vert F^{\dagger }\left\vert \psi
\right\rangle |^{2}=\pi \int d^{2}\gamma d^{2}\sigma \delta \left( \eta
_{2}-D\sigma _{2}+B\gamma _{1}\right) \delta \left( \eta _{1}-D\sigma
_{1}-B\gamma _{2}\right) \left\langle \psi \right\vert \Delta \left( \sigma
,\gamma \right) \left\vert \psi \right\rangle .  \label{36}
\end{equation}%
where $\left\langle \psi \right\vert \Delta \left( \sigma ,\gamma \right)
\left\vert \psi \right\rangle $ is the Wigner function. So the probability
distribution for the Fresnel quadrature phase \textbf{(}see Eq. (A11) in the
Appendix) is the tomography (Radon transform of the two-mode Wigner
function).{\small \ }This is the main result of the present paper. This new
relation between quantum tomography and optical Fresnel transform may
provide experimentalists to figure out new approach for generating
tomography.

\section{In the conjugate representation}

Next we turn to the \textquotedblleft frequency\textquotedblright\ domain,
that is to say, we shall prove that the $(A,C)$ related Radon transform of
entangled Wigner operator $\Delta \left( \sigma ,\gamma \right) $ is just
the pure state density operator $\left\vert \xi \right\rangle
_{s,rs,r}\left\langle \xi \right\vert ,$ i.e.,
\begin{equation}
F_{2}\left\vert \xi \right\rangle \left\langle \xi \right\vert
F_{2}^{\dagger }=\left\vert \xi \right\rangle _{s,rs,r}\left\langle \xi
\right\vert =\pi \int \delta \left( \xi _{1}-A\sigma _{1}-C\gamma
_{2}\right) \delta \left( \xi _{2}-A\sigma _{2}+C\gamma _{1}\right) \Delta
\left( \sigma ,\gamma \right) d^{2}\sigma d^{2}\gamma ,  \label{37}
\end{equation}%
where%
\begin{equation}
\left\vert \xi \right\rangle =\exp \left[ -\frac{1}{2}\left\vert \xi
\right\vert ^{2}+\xi a_{1}^{\dagger }+\xi ^{\ast }a_{2}^{\dagger
}-a_{1}^{\dagger }a_{2}^{\dagger }\right] \left\vert 00\right\rangle
\label{38}
\end{equation}%
is an entangled state conjugate to $\left\vert \eta \right\rangle .$ By
analogy with the above procedure, we obtain the 2-dimensional Fresnel
transformation in its `frequency domain', i.e.,
\begin{eqnarray}
\mathcal{K}_{2}^{N}\left( \xi ^{\prime },\xi \right)  &\equiv &\frac{1}{\pi }%
\left\langle \xi ^{\prime }\right\vert F_{2}\left( r,s\right) \left\vert \xi
\right\rangle   \notag \\
&=&\int \frac{d^{2}\eta d^{2}\sigma }{\pi ^{2}}\left\langle \xi ^{\prime
}\right\vert \left. \eta ^{\prime }\right\rangle \left\langle \eta ^{\prime
}\right\vert F_{2}\left( r,s\right) \left\vert \eta \right\rangle
\left\langle \eta \right\vert \left. \xi \right\rangle   \notag \\
&=&\frac{1}{8iB\pi }\int \frac{d^{2}\sigma d^{2}\eta }{\pi ^{2}}\exp \left(
\frac{\xi ^{\prime \ast }\eta ^{\prime }-\xi ^{\prime }\eta ^{\prime \ast
}+\xi \eta ^{\ast }-\xi ^{\ast }\eta }{2}\right) \mathcal{K}_{2}^{\left(
\mathtt{r},s\right) }\left( \sigma ,\eta \right)   \notag \\
&=&\frac{1}{2i\left( -C\right) \pi }\exp \left[ \frac{i}{2\left( -C\right) }%
\left( D\left\vert \xi \right\vert ^{2}+A\left\vert \xi ^{\prime
}\right\vert ^{2}-\xi ^{\prime \ast }\xi -\xi ^{\prime }\xi ^{\ast }\right) %
\right] ,  \label{39}
\end{eqnarray}%
where the superscript $N$ means that this transform kernel corresponds to
the parameter matrix $N=\left[ D,-C,-B,A\right] $. Thus the 2D Fresnel
transformation in its `frequency domain' is given by%
\begin{equation}
\Psi \left( \xi ^{\prime }\right) =\int \mathcal{K}_{2}^{N}\left( \xi
^{\prime },\xi \right) \Phi \left( \xi \right) d^{2}\xi .  \label{40}
\end{equation}%
Operating $F_{2}\left( r,s\right) $ on $\left\vert \xi \right\rangle $ we
have (also see Appendix)%
\begin{equation}
\left\vert \xi \right\rangle _{s,r}=\frac{1}{\allowbreak \allowbreak A-iC}%
\exp \left\{ -\frac{D+iB}{2\left( \allowbreak A-iC\right) }\left\vert \eta
\right\vert ^{2}+\frac{\xi a_{1}^{\dagger }}{A-iC}+\frac{\xi ^{\ast
}a_{2}^{\dagger }}{\allowbreak A-iC}-\frac{\allowbreak A+iC}{\allowbreak A-iC%
}a_{1}^{\dagger }a_{2}^{\dagger }\right\} \left\vert 00\right\rangle ,
\label{41}
\end{equation}%
or
\begin{equation}
\left\vert \xi \right\rangle _{s,r}=\frac{1}{s^{\ast }-r^{\ast }}\exp
\left\{ -\allowbreak \frac{s^{\ast }+r^{\ast }}{2\left( s^{\ast }-r^{\ast
}\right) }\left\vert \xi \right\vert ^{2}+\allowbreak \frac{\xi
a_{1}^{\dagger }}{s^{\ast }-r^{\ast }}\allowbreak +\allowbreak \frac{\xi
^{\ast }a_{2}^{\dagger }}{s^{\ast }-r^{\ast }}-\frac{s-r}{s^{\ast }-r^{\ast }%
}\allowbreak a_{1}^{\dagger }a_{2}^{\dagger }\right\} \left\vert
00\right\rangle .  \label{42}
\end{equation}%
Noticing that the entangled Wigner operator in $\left\langle \xi \right\vert
$ representation is expressed as
\begin{equation}
\Delta \left( \sigma ,\gamma \right) =\int \frac{d^{2}\xi }{\pi ^{3}}%
\left\vert \gamma +\xi \right\rangle \left\langle \gamma -\xi \right\vert
\exp (\xi ^{\ast }\sigma -\sigma ^{\ast }\xi ),  \label{43}
\end{equation}%
and using the classical correspondence of $\left\vert \xi \right\rangle
_{s,rs,r}\left\langle \xi \right\vert $ which is calculated by
\begin{eqnarray}
h(\sigma ,\gamma ) &=&4\pi ^{2}\mathtt{Tr}\left[ \left\vert \xi
\right\rangle _{s,r\text{ }s,r}\left\langle \xi \right\vert \Delta \left(
\sigma ,\gamma \right) \right]   \notag \\
\  &=&4\int \frac{d^{2}\xi }{\pi }\left\langle \gamma -\xi \right\vert
F_{2}\left\vert \xi \right\rangle \left\langle \xi \right\vert F_{2}^{\dag
}|\gamma +\xi \rangle \exp (\xi ^{\ast }\sigma -\sigma ^{\ast }\xi )  \notag
\\
&=&\pi \delta \left( \xi _{1}-A\sigma _{1}-C\gamma _{2}\right) \delta \left(
\xi _{2}-A\sigma _{2}+C\gamma _{1}\right) ,  \label{44}
\end{eqnarray}%
we obtain
\begin{equation}
\left\vert \xi \right\rangle _{s,r\text{ }s,r}\left\langle \xi \right\vert
=\pi \int \delta \left( \xi _{1}-A\sigma _{1}-C\gamma _{2}\right) \delta
\left( \xi _{2}-A\sigma _{2}+C\gamma _{1}\right) \Delta \left( \sigma
,\gamma \right) d^{2}\sigma d^{2}\gamma ,  \label{45}
\end{equation}%
so the projector operator $\left\vert \xi \right\rangle _{s,r\text{ }%
s,r}\left\langle \xi \right\vert $ is another Radon transformation of the
two-mode Wigner operator, with $A$ and $C$ being the Radon transformation
parameter (`frequency' domain). Therefore, the quantum tomography in $%
_{s,r}\left\langle \xi \right\vert $ representation is expressed as the
Radon transformation of the Wigner function%
\begin{equation}
|\left\langle \xi \right\vert F^{\dagger }\left\vert \psi \right\rangle
|^{2}=|_{s,r}\left\langle \xi \right\vert \left. \psi \right\rangle
|^{2}=\pi \int d^{2}\gamma d^{2}\sigma \delta \left( \xi _{1}-A\sigma
_{1}-C\gamma _{2}\right) \delta \left( \xi _{2}-A\sigma _{2}+C\gamma
_{1}\right) \left\langle \psi \right\vert \Delta \left( \sigma ,\gamma
\right) \left\vert \psi \right\rangle ,  \label{46}
\end{equation}%
and $_{s,r}\left\langle \xi \right\vert =\left\langle \xi \right\vert
F^{\dagger }.$

\section{Inverse Radon transformation}

Now we consider the inverse Radon transformation. For instance, using (\ref%
{35}) we see the Fourier transformation of $\left\vert \eta \right\rangle
_{s,rs,r}\left\langle \eta \right\vert $ is%
\begin{eqnarray}
&&\int d^{2}\eta \left\vert \eta \right\rangle _{s,rs,r}\left\langle \eta
\right\vert \exp (-i\zeta _{1}\eta _{1}-i\zeta _{2}\eta _{2})  \notag \\
&=&\pi \int d^{2}\gamma d^{2}\sigma \Delta \left( \sigma ,\gamma \right)
\exp \left[ -i\zeta _{1}\left( D\sigma _{1}+B\gamma _{2}\right) -i\zeta
_{2}\left( D\sigma _{2}-B\gamma _{1}\right) \right] ,  \label{47}
\end{eqnarray}%
the right-hand side of (\ref{47}) can be regarded as a special Fourier
transformation of $\Delta \left( \sigma ,\gamma \right) $, so by making its
inverse Fourier transformation, we get%
\begin{eqnarray}
\Delta \left( \sigma ,\gamma \right) &=&\frac{1}{(2\pi )^{4}}\int_{-\infty
}^{\infty }dr_{1}\left\vert r_{1}\right\vert \int_{-\infty }^{\infty
}dr_{2}\left\vert r_{2}\right\vert \int_{0}^{\pi }d\theta _{1}d\theta _{2}
\notag \\
&&\times \int_{-\infty }^{\infty }\frac{d^{2}\eta }{\pi }\left\vert \eta
\right\rangle _{s,rs,r}\left\langle \eta \right\vert K\left(
r_{1},r_{2},\theta _{1},\theta _{2}\right) ,  \label{48}
\end{eqnarray}%
where $\cos \theta _{1}=\cos \theta _{2}=\frac{D}{\sqrt{B^{2}+D^{2}}}%
,r_{1}=\zeta _{1}\sqrt{B^{2}+D^{2}},r_{2}=\zeta _{2}\sqrt{B^{2}+D^{2}}$ and
\begin{eqnarray}
K\left( r_{1},r_{2},\theta _{1},\theta _{2}\right) &\equiv &\exp \left[
-ir_{1}\left( \frac{\eta _{1}}{\sqrt{B^{2}+D^{2}}}-\sigma _{1}\cos \theta
_{1}-\gamma _{2}\sin \theta _{1}\right) \right]  \notag \\
&&\times \exp \left[ -ir_{2}\left( \frac{\eta _{2}}{\sqrt{B^{2}+D^{2}}}%
-\sigma _{2}\cos \theta _{2}+\gamma _{1}\sin \theta _{2}\right) \right] .
\label{49}
\end{eqnarray}%
Eq.(\ref{48}) is just the inverse Radon transformation of entangled Wigner
operator in the entangled state representation. This is different from the
two independent Radon transformations' direct product of the two independent
single-mode Wigner operators, because in (\ref{23}) the $\left\vert \eta
\right\rangle _{s,r}$ is an entangled state. Therefore the Wigner function
of quantum state $\left\vert \psi \right\rangle $ can be reconstructed from
the tomographic inversion of a set of measured probability distributions $%
\left\vert _{s,r}\left\langle \eta \right. \left\vert \psi \right\rangle
\right\vert ^{2}$, i.e.,%
\begin{eqnarray}
W_{\psi } &=&\frac{1}{(2\pi )^{4}}\int_{-\infty }^{\infty }dr_{1}\left\vert
r_{1}\right\vert \int_{-\infty }^{\infty }dr_{2}\left\vert r_{2}\right\vert
\int_{0}^{\pi }d\theta _{1}d\theta _{2}  \notag \\
&&\times \int_{-\infty }^{\infty }\frac{d^{2}\eta }{\pi }\left\vert
_{s,r}\left\langle \eta \right. \left\vert \psi \right\rangle \right\vert
^{2}K\left( r_{1},r_{2},\theta _{1},\theta _{2}\right) .  \label{50}
\end{eqnarray}

\bigskip

In summary, based on the preceding paper \cite{r10}, we have further
extended the relation connecting optical Fresnel transformation with
quantum
tomography to the entangled case{\small .} The tomography representation $%
_{s,r}\left\langle \eta \right\vert =\left\langle \eta \right\vert
F_{2}^{\dagger }$ is set up, based on which the tomogram of quantum state $%
\left\vert \psi \right\rangle $ is just the squared modulus of the wave
function $_{s,r}\left\langle \eta \right\vert \left. \psi \right\rangle .$
i.e. the probability distribution for the Fresnel quadrature phase is the
tomogram (Radon transform of the Wigner function).

\textbf{Acknowledgement: }Work supported by the National Natural Science
Foundation of China (Grant No 10874174) and the President Foundation of
Chinese Academy of Science, and the Research Foundation of the Education
Department of Jiangxi Province.

\textbf{APPENDIX}

In fact, from (\ref{13}) we have
\begin{align}
F_{2}Q_{1}F_{2}^{\dagger }& =\frac{1}{2}\left( \left( A+D\right)
Q_{1}-\left( B-C\right) P_{1}+\left( A-D\right) Q_{2}+\left( B+C\right)
P_{2}\right) ,  \tag{A1} \\
F_{2}Q_{2}F_{2}^{\dagger }& =\frac{1}{2}\left( \left( A+D\right)
Q_{2}-\left( B-C\right) P_{2}+\left( A-D\right) Q_{1}+\left( B+C\right)
P_{1}\right) ,  \tag{A2}
\end{align}%
and
\begin{align}
F_{2}P_{1}F_{2}^{\dagger }& =\frac{1}{2}\left( \left( A+D\right)
P_{1}+\left( B-C\right) Q_{1}-\left( A-D\right) P_{2}+\left( B+C\right)
Q_{2}\right) ,  \tag{A3} \\
F_{2}P_{2}F_{2}^{\dagger }& =\frac{1}{2}\left( \left( A+D\right)
P_{2}+\left( B-C\right) Q_{2}-\left( A-D\right) P_{1}+\left( B+C\right)
Q_{1}\right) ,  \tag{A4}
\end{align}%
it then follow that
\begin{align}
F_{2}\left( Q_{1}-Q_{2}\right) F_{2}^{\dagger }& =D\left( Q_{1}-Q_{2}\right)
-B\left( P_{1}-P_{2}\right) ,  \tag{A5} \\
F_{2}\left( P_{1}+P_{2}\right) F_{2}^{\dagger }& =B\left( Q_{1}+Q_{2}\right)
+D\left( P_{1}+P_{2}\right) ,  \tag{A6}
\end{align}%
and
\begin{align}
F_{2}\left( Q_{1}+Q_{2}\right) F_{2}^{\dagger }& =A\left( Q_{1}+Q_{2}\right)
+C\left( P_{1}+P_{2}\right) ,  \tag{A7} \\
F_{2}\left( P_{1}-P_{2}\right) F_{2}^{\dagger }& =A\left( P_{1}-P_{2}\right)
-C\left( Q_{1}-Q_{2}\right) .  \tag{A8}
\end{align}%
Noticing that $\left[ F_{2}\left( Q_{1}-Q_{2}\right) F_{2}^{\dagger
},F_{2}\left( P_{1}+P_{2}\right) F_{2}^{\dagger }\right] =0$ and (\ref{17})
thus the eigenvector equation of communicative operators $D\left(
Q_{1}-Q_{2}\right) -B\left( P_{1}-P_{2}\right) $ and $B\left(
Q_{1}+Q_{2}\right) +D\left( P_{1}+P_{2}\right) $ is
\begin{align}
\left[ D\left( Q_{1}-Q_{2}\right) -B\left( P_{1}-P_{2}\right) \right]
\left\vert \eta \right\rangle _{s,r}& =F_{2}\left( Q_{1}-Q_{2}\right)
F_{2}^{\dagger }\left\vert \eta \right\rangle _{s,r}=\sqrt{2}\eta
_{1}\left\vert \eta \right\rangle _{s,r},\text{ }  \tag{A9} \\
\left[ B\left( Q_{1}+Q_{2}\right) +D\left( P_{1}+P_{2}\right) \right]
\left\vert \eta \right\rangle _{s,r}& =F_{2}\left( P_{1}+P_{2}\right)
F_{2}^{\dagger }\left\vert \eta \right\rangle _{s,r}=\sqrt{2}\eta
_{2}\left\vert \eta \right\rangle _{s,r},  \tag{A10}
\end{align}%
thus
\begin{equation}
\left\vert \eta \right\rangle _{s,r}=F_{2}\left\vert \eta \right\rangle =%
\text{Eq.(\ref{22})},  \tag{A11}
\end{equation}%
and we name $D\left( Q_{1}-Q_{2}\right) -B\left( P_{1}-P_{2}\right) $\ or $%
B\left( Q_{1}+Q_{2}\right) +D\left( P_{1}+P_{2}\right) $\ the Fresnel
quadrature phase.

On the other hand, due to the communicative relation $\left[ F_{2}\left(
Q_{1}+Q_{2}\right) F_{2}^{\dagger },F_{2}\left( P_{1}-P_{2}\right)
F_{2}^{\dagger }\right] =0,$ and $\left\vert \xi \right\rangle $ (the
conjugate state to $\left\vert \eta \right\rangle $) is the common
eigen-equation of $\left( Q_{1}+Q_{2}\right) $ and $\left(
P_{1}-P_{2}\right) $, i.e.,
\begin{equation}
\left( Q_{1}+Q_{2}\right) \left\vert \xi \right\rangle =\sqrt{2}\xi
_{1}\left\vert \xi \right\rangle ,\text{ }\left( P_{1}-P_{2}\right)
\left\vert \xi \right\rangle =\sqrt{2}\xi _{2}\left\vert \xi \right\rangle ,
\tag{A12}
\end{equation}%
so the common eigenvector of $F_{2}\left( Q_{1}+Q_{2}\right) F_{2}^{\dagger
} $ and $F_{2}\left( P_{1}-P_{2}\right) F_{2}^{\dagger }$ is given by%
\begin{align}
\left\vert \xi \right\rangle _{s,r}& \equiv F_{2}\left\vert \xi
\right\rangle =F_{2}\int \frac{d^{2}\eta }{\pi }\left\vert \eta
\right\rangle \left\langle \eta \right\vert \left. \xi \right\rangle  \notag
\\
& =\int \frac{d^{2}\eta }{2\pi }\exp \left( \frac{\xi \eta ^{\ast }-\xi
^{\ast }\eta }{2}\right) \left\vert \eta \right\rangle _{s,r}  \notag \\
& =\text{Eq.(\ref{41})=Eq.(\ref{42}),}  \tag{A13}
\end{align}%
where we have used the overlap relation of $\left\langle \eta \right\vert
\left. \xi \right\rangle =\frac{1}{2}\exp \left( \frac{\xi \eta ^{\ast }-\xi
^{\ast }\eta }{2}\right) $. The corresponding eigen-equations of $\left\vert
\xi \right\rangle _{s,r}$ are%
\begin{align}
\left[ A\left( Q_{1}+Q_{2}\right) +C\left( P_{1}+P_{2}\right) \right]
\left\vert \xi \right\rangle _{s,r}& =\sqrt{2}\xi _{1}\left\vert \xi
\right\rangle _{s,r},  \tag{A14} \\
\left[ A\left( P_{1}-P_{2}\right) -C\left( Q_{1}-Q_{2}\right) \right]
\left\vert \xi \right\rangle _{s,r}& =\sqrt{2}\xi _{2}\left\vert \xi
\right\rangle _{s,r}.  \tag{A15}
\end{align}

\

\end{document}